\documentstyle[11pt]{article}
\begin{document}
\parindent=5pt
\topmargin=0 cm
\hsize37truepc\vsize61truepc
\hoffset=-.5truein\voffset=-0.8truein
\thispagestyle{empty}
\baselineskip= 16pt plus 1pt minus 1pt
\textheight= 23cm
\textwidth= 15cm
\overfullrule=0pt
\parskip 3 pt plus 1pt minus 1 pt
\def\p{\partial} \def\D{{\cal D}}
\def\bD{{\bar {\cal D}}}
\def\Z{\Xi}
\def\e{\epsilon} \def\g{\gamma}
\def\n{\nabla}\def\hn{{\widehat\nabla}} \def\d{\delta}
\def\r{\rho} \def\s{\sigma} \def\z{\zeta}
\def\eb{{\bar \eta}} \def\x{\chi}
\def\t{\vartheta} \def\bt{{\bar \vartheta}}
\def\b{\beta} \def\a{\alpha} \def\l{\lambda}
\def\m{\mu} \def\f{\varphi}
\def\da{{\dot\alpha}} \def\db{{\dot\beta}}
\def\dg{{\dot\gamma}}
\def\dd{{\dot\delta}} \def\de{{\dot\eta}}
\def\op{\oplus} \def\ot{\otimes}
\def\eom{equation of motion} \def\eoms{equations of motion}
\def\eq{equation}
\def\hs{harmonic space } \def\h{harmonic }
\def\ji{Jacobi identity}\def\jis{Jacobi identities}
\def\lc{leading component} \def\rep{representation}
\def\sdym{self-dual Yang-Mills }
\def\sdyme{self-dual Yang-Mills equations }
\def\sd{self-dual }\def\sdy{self-duality }
\def\ssdy{super self-duality }\def\ssd{super self-dual }
\def\sf{superfield} \def\sp{super-Poincar\'e algebra}
\def\sym{super Yang-Mills } \def\ym{Yang-Mills }
\def\susy{supersymmetric }\def\ss{superspace }
\def\sg{supergravity } \def\te{$\bt$-expansion}
\def\half{{1\over 2}} \def\N#1{$N=#1$ }
\def\dim#1{dimension $#1$ }
\def\Tr{\mbox {Tr }}\def\M{${\cal M}^4$}
\def\sM{${\cal M}^{4|2N}$}
\def\fn{\phi_{i_{n+2}\dots i_1 \da_1\dots \da_{n}}}
\def\gn{\phi_{i_{n+3}\dots i_1 \da_1\dots \da_{n+1}}}
\def\0#1{{\stackrel{\circ}{#1}}} \def\L{{\cal L}}
\def\der#1{{\partial \over \partial #1}}
\def\pp#1#2{{\partial #1 \over \partial #2}}
\def\be{\begin{equation}}
\def\r#1{(\ref{#1})} \def\la#1{\label{#1}}
\def\c#1{\cite{#1}}
\def\ee{\end{equation}}
\def\arr{\begin{array}{rll}}
\def\ea{\end{array}}
\def\bea{\begin{eqnarray}}
\def\eea{\end{eqnarray}}
\def\bealph#1
{\setcounter{equation}{0}\renewcommand\theequation{#1\alph{equation}}\bea}
\def\eealph#1
{\eea\setcounter{equation}{#1}\renewcommand\theequation{\arabic{equation}}}
\def\square{\kern1pt\vbox
            {\hrule height 0.6pt\hbox{\vrule width 0.6pt\hskip 3pt
  \vbox{\vskip 6pt}\hskip 3pt\vrule width 0.6pt}\hrule height 0.6pt}\kern1pt}
\rightline{hep-th/9606027}
\rightline{IC/96/88}
\vskip 1 true cm
\centerline{{\Large         Interacting Fields of Arbitrary Spin         }}
\vskip .1 true cm
\centerline{{\Large                        and                           }}
\vskip .1 true cm
\centerline{{\Large   $N>4$ Supersymmetric Self-dual Yang-Mills Equations}}
\vskip 1.0 true cm
\centerline{{\large             Ch. Devchand$^1$  and  V. Ogievetsky  }}
\vskip .5 true cm{\small
\centerline{{\it     Joint Institute for Nuclear Research, Dubna, Russia}}
\centerline{{\it  and      }}
\centerline{{\it $^1$International Centre for Theoretical Physics, 
                                  Trieste, Italy     }}}
\vskip 1 true cm
\noindent
{\bf Abstract.}
We show that the self-dual Yang-Mills equations afford supersymmetrisation
to systems of equations invariant under global $N$-extended super-Poincar\'e
transformations for arbitrary values of $N$, without the limitation ($N\le 4$)
applicable to standard non-self-dual Yang-Mills theories. These systems of
equations provide novel classically consistent interactions for vector
supermultiplets containing fields of spin up to ${N-2 \over 2}$.
The equations of motion for the component fields of spin greater than 
$\half$ are interacting variants of the first-order Dirac--Fierz equations 
for zero rest-mass fields of arbitrary spin. The interactions are governed 
by conserved currents which are constructed by an iterative procedure.
In (arbitrarily extended) chiral superspace, the equations of motion for 
the (arbitrarily large) self-dual supermultiplet are shown to be completely
equivalent to the set of algebraic supercurvature constraints defining
the self-dual superconnection.
\vskip 10pt
\section{Introduction}
\vskip 10pt
{\bf 1.1 } It is widely believed that massless fields of spin greater
than two cannot be consistently coupled to lower spin fields. This belief
is based on the paucity of appropriate conserved charges.
In the case of the free Maxwell equations for a set of spin
one fields, $ \p^\mu F^a_{\mu\nu}=0, a=1...n$, conceivable interactions
are introduced via source currents,
\be  \p^\mu F^a_{\mu\nu} = j_\nu^a\  . \la{M}\ee
Consistency then requires the conservation of these currents,
\be \p^\nu j^a_\nu = 0\  ,\la{C}\ee since $\p^\mu \p^\nu F^a_{\mu\nu} = 0 $
in virtue of the antisymmetry of $F^a_{\mu\nu}$. These conservation laws
are related to the gauge invariance of the Maxwell equations \c{op1}.
Further analysis shows that for massless vector fields the only consistent
equations for interacting spin--one fields without higher derivatives are
the Yang-Mills equations \c{op2}. Similarly, for a massless symmetric tensor
field \c{fp}, the only classically consistent theory \c{op3} is the Einstein
theory of gravity, with a conserved stress tensor as the source of
interactions. Consistent coupling of massless spin 2 fields with spin
${3\over 2}$ fields yields the supergravity equations,
with conserved spin-vector currents as sources of gravitino fields.
There are not many examples of classically consistent relativistic
equations for interacting massless fields of spin greater than two.
Consistent light-cone frame interacting theories of higher spin fields
with arbitrarily extended supersymmetry do however exist \c{bbb}. Furthermore,
consistent systems of infinitely many fields of every (half-) interger
spin are known from string theories and otherwise (e.g. \c{v}).

We shall show in the present paper that the possibilities for coupling
higher spin fields to lower spin ones are dramatically changed in spaces 
of signature (4,0) or (2,2) (or in complex space) provided that the gauge
field is taken to be self-dual. In the theory of interactions amongst higher
spin fields, this is a hitherto unexplored possibility, which yields the
unexpected result that the higher spin fields satisfy classically 
consistent interacting forms of the first-order zero rest-mass  
Dirac--Fierz equations \c{d,f}. 
The absence of conjugation between dotted and undotted spinor indices in 
these spaces weakens the compatibility conditions, which in Minkowski space
turn out to be forbiddingly strong \c{f,fp}.

\vskip 10pt
{\bf 1.2 }
In two-spinor language, with dotted and undotted indices raised and lowered
by the skew-symmetric symplectic invariants $\e_{\a\b}, \e_{\da\db},
\e^{\a\b}, \e^{\da\db} $, Maxwell's equations take the form
\be \e^{\da\dg} \p_{\a\dg} f_{\da\db} + \e^{\b\g} \p_{\g\db} f_{\a\b}
    = 2 j_{\a\db}\   ,\la{m}\ee
where the symmetric tensor $f_{\da\db}$ describes the helicity $+1$ component
and its Minkowski space conjugate  $f_{\a\b}$ the helicity $-1$
component of the Maxwell field. The Bianchi identity
$\p_{[\mu} F_{\nu\rho ]} =0 $ however takes the form
\be \e^{\da\dg} \p_{\a\dg} f_{\da\db} - \e^{\b\g} \p_{\g\db} f_{\a\b} =0 ,\ee
so the complete Maxwell equations may be compactly written in either the form
\be \p_\a^{~\dg} f_{\dg\db} = j_{\a\db}  \la{m1}\ee
or the conjugate form
\be \p^\b_{~\db} f_{\a\b} = j_{\a\db}\  ,\la{m2}\ee
provided the current is real.
Generalising \r{m1}, a massless field of helicity $s$ may be described by a 
symmetric rank $n=2s$ spinor $\f_{\da_1 \dots \da_{n}}$ satisfying
\be \p_\a^{\da_n} \f_{\da_1 \dots \da_{n}} =
                         j_{\a\da_1 \dots \da_{n-1}}\ .\la{a}\ee
Now differentiating \r{a} and using the identity
$\p^{\a\da}\p^{\db}_\a = \half \e^{\da\db} \p^\a_{\dg} \p^{\dg}_\a $,
we see that antisymmetry of $\e$ and symmetry of $\f$ in the dotted indices
requires for consistency, that the current on the right is divergence free,
viz. \be     \p^{\a\da_1} j_{\a\da_1 \dots \da_{n-1}} = 0\   .\la{C1} \ee
This generalises the argument for current conservation \r{C}
in the Maxwell case above.
In fact the investigation of source--free versions of \r{a}, 
the `zero rest-mass' equations of Dirac and Fierz,
\be   \p_\g^{\da} \f_{\da\db\dots \dd }  =\ 0\   ,\la{free}\ee
has a long history (see e.g. \c{d,f,p}). In particular, Fierz \c{f}
discussed the problems of consistently coupling such fields to an external
electromagnetic field. He noticed that the minimal substitution
of $\p_{\a\da}$ in \r{free} by the gauge covariant derivative
$\n_{\a\da} = \p_{\a\da} + A_{\a\da}$ requires the
satisfaction of $f_{\da\db}=0$ for consistency. This allows only a
pure--gauge spin 1 coupling in Minkowski space, where
fields with dotted and undotted indices are related by complex conjugation,
and eq. \r{a} is equivalent to the conjugate equation for the helicity +s
field,
\be \p_\da^{\a_n} \f_{\a_1 \dots \a_{n}} =
                        j_{\da\a_1 \dots \a_{n}}\ .\la{b}\ee
However, in spaces of signature (4,0) or (2,2) the conjugation between
dotted and undotted spinors is lifted, opening new horizons.
Eqs. \r{a} and \r{b} can then be considered independently. In particular,
there is then actually no problem in consistently coupling an external
{\it \sd} gauge field to such spinors, for the requirement
$f_{\da\db}=0$ is precisely the self-duality equation.
Now, if a non-zero current $J$ is present on the right-hand side of \r{free},
consistency of such a minimal coupling, i.e. of the first--order equation
\be   \n_\g^{\da} \f_{\da\db \dots \dd} =
                 J_{\g\db \dots \dd}\   ,\la{int}\ee
further requires the covariant constancy of this current,
\be     \n^{\g\db}J_{\g\db\dots \dd} = 0\   .\la{ccc}\ee
Remarkably, such conserved currents are provided by the higher-N
supersymmetric couplings of \sdym fields.

\vskip 10pt
{\bf 1.3 }
It is noteworthy that higher-spin fields satisfying the interacting 
Dirac--Fierz equations \r{int} are precisely the ones which are needed
in order to supersymmetrise the \sdyme beyond the traditionally `maximal' 
\N4 extension. Consistent lower-spin \eoms\ recursively give rise to 
consistent higher-spin \eoms, so that the \sdy \eq s  may be extended to 
systems of equations invariant under the N--extended \sp\ for {\it any} 
choice of $N$. The self-dual vector supermultiplet may therefore be made
as large as one desires, to contain a spectrum of fields up to any given
spin ${N-2\over 2}$. 
Classical consistency of the equations of motion does not set any limit on 
the extension $N$. This is unlike the situation for the full non-\sd \ym
equations which have \N4 as the `maximal' extension \c{bss}. We have already
announced the existence of these higher--N supersymmetric systems in \c1,
where we explicitly displayed the \eoms\  for the first two unconventional
extensions, \N5 and \N6. The purpose of this paper is to provide the complete
proof of the consistency of these \ssdy \eq s for arbitrarily large $N$ in a
manifestly supersymmetric superspace setting. These \ssd systems are
manifestly covariant four dimensional globally supersymmetric realisations
of $N>4$ \sp s. They also provide classically consistent manifestly
covariant \eoms\  for interacting massless fields of spin greater than two,
which are moreover possibly the only consistently coupled
generalisations of the zero-rest mass Dirac-Fierz equations \r{free}.

Up to and including \N3 the \ssd \ym systems \c{matr} are reductions
of the corresponding full non-\sd \sym \eq s. The former are constrained 
versions of the latter, with
spectra consisting of precisely half the fields. For \N3, for instance,
the spectrum  of the full \sym theory consists of the following two
irreducible \rep s of the \sp\
$$ \{ f_{\a\b} ,\quad \l_{i\a},\quad W_{ij},\quad \x_{\da} \}\qquad
   \mbox{and} \qquad
\{ f_{\da\db} ,\quad \l^i_\da ,\quad \overline W^{ij},\quad \x_\a \},$$
where the $f_{\a\b}$ (resp. $f_{\da\db}$) are the (anti-) \sd parts of
the \ym field-strength
$$ [\n_{\a \da}, \n_{\b \db}] :=
   \e_{\da \db} f_{\a \b} + \e_{\a \b} f_{\da \db}\   ,$$
the $\l$'s and $\x$'s are spin one-half fields, and the $W$'s are scalar
fields; all fields taking values in the Lie algebra of the gauge group.
In Minkowski space these two supermultiplets are conjugate to each
other, but in spaces of signature (4,0) or (2,2), or complexified space,
they are independent; and the \ssdy restriction
is precisely the condition that the anti-\sd right-hand multiplet above is
zero. Constraining the full \sym \eq s in this fashion yields the \ssdy 
\eq s. Thus the equations of motion for the full N=3 theory \c{hhls}
$$\arr
\e^{\da\dg} \n_{\a\dg} f_{\da\db} &+& \e^{\b\g} \n_{\g\db} f_{\a\b}
=\ \{\l_{\a i}, \l_\db^i \}  + \{\x_{\a }, \x_\db \} +
 [ W_i ,\n_{\a\db} W^i ] + [W^i  ,\n_{\a\db} W_i ]\\[5pt]
&& \e^{\dg\da} \n_{\a\dg} \l_\da^i\ = - \e^{ijk} [\l_{j\a} , W_k]
				     + [\x_{\a }, W^i ]   \\[5pt]
&& \e^{\g\b} \n_{\g\db} \l_{i\b}\ = - \e_{ijk} [\l_{\db}^j , W^k]
				   + [\x_{\db }, W_i ] \\[5pt]
&& \e^{\dg\da} \n_{\a\dg} \x_{\da }\ = - [\l_{k\a} , W^k]  \\[5pt]
&& \e^{\g\b} \n_{\g\db} \x_{\b}\ = - [\l_{\db}^k , W_k] \\[5pt]
\n_{\a\db}\n^{\a\db} W^i & =& - 2 [[W_j , W^i], W^j] + [[W_j , W^j], W^i]
  +\half \e^{ijk}\{\l^\a_j , \l_{k\a} \} + \{\l^{i\da} , \x_\da \}   \\[5pt]
\n_{\a\db}\n^{\a\db} W_i & =& - 2 [[W^j , W_i], W_j] + [[W^j , W_j], W_i]
  +\half \e_{ijk} \{\l^{j\da}, \l_{\da}^k \} + \{\l^{\a}_i , \x_\a \}
\ea$$
reduce to the \N3 \ssdy equations, which we write in $N$-independent fashion
using $\x_{ijk\da} = \e_{ijk}\x_\da$ and $W_{ij} = \e_{ijk}W^k$,
\be\arr
\e^{\b\g} \n_{\g\db} f_{\a\b} &=& 0\\[5pt]
 \e^{\g\b} \n_{\g\db} \l_{i\b} &=& 0 \\[5pt]
\e^{\da\dg} \n_{\a\dg} \x_{ijk\da } &=& [\l_{[i\a} , W_{jk]}]  \\[5pt]
 \n_{\a\db}\n^{\a\db} W_{ij}  &=&  \{\l_i^{\a}, \l_{j\a} \}.
\la{n3sd}\ea\ee
The first equation is identically satisfied in virtue of the Bianchi
identity and the condition $ f_{\da\db} = 0$ which is just the usual
(N=0) first order \sdy equation for the vector potential,
$ F_{\mu\nu} = \half\e_{\mu\nu\rho\s} F^{\rho\s}$.
For \N4, however, there is no similar correspondence between the standard
maximally supersymmetric \sym \eq s \c{bss} and the \N4 supersymmetrisation
of the \sdy condition $f_{\da\db}=0$ \c{s}. Being irreducible, the
standard \N4 Yang-Mills multiplet
$$ \{ f_{\a\b} ,\quad \l_{i\a},\quad W_{ij},\quad \x^i_{\da},\quad 
f_{\da\db} \} $$ does not admit the the self-duality constraint 
of the above type. In this case the so-called reality constraint 
for the scalar \sf\ $W_{ij} = \half \e_{ijkl} \overline W^{kl}$, which 
imposes Minkowski space self-conjugacy of the \N4 multiplet, needs to be
lifted. Thus having doubled the representation, the self-duality constraint 
$\overline W^{kl} = 0$ may now be imposed instead of the reality constraint
in order to reduce the multiplet to the irreducible \sd one \c{s}. 
Therefore, not only are the \N4 \ssdy \eq s  not restrictions of the full
non-\sd \eq s, but the spectrum of the former is in no sense a restriction
of the latter and for $N>4$ there do not even exist non-\sd \eq s 
corresponding to the $N>4$ extended \ssdy \eq s  which we present in this 
paper. For the \N4 case, the above \N3 \eq s remain unchanged and are merely
enhanced by an \eq\ of the form  
\be \e^{\da\db} \n_{\a\db} g_{ijkl\da\dg } =\  J_{ijkl \a\dg}\  ,\la{n4}\ee
for the additional spin 1 field $ g_{ijkl\da\dg}$, where the current on 
the right satisfies \r{ccc} in virtue of the lower spin equations \r{n3sd}.
This pattern actually repeats itself for higher spin fields, yielding
first-order equations of the form
\be \n_\a^{\da_n} \phi_{i_{n+2}\dots i_1 \da_1\dots \da_{n} }  = 
            J_{i_{n+2}\dots i_1 \a\da_1\dots \da_{n-1} }\   ,\ee 
for arbitrary $n\ge 2$ up to $n=N-2$, essentially because the 
$(N-1)$--extended system, which contains fields of spin up to 
${(N-3)\over 2}$, nestles within the $N$--extended system 
completely intact, and provides a conserved source current for a 
new spin ${(N-2)\over 2}$ field. 
The $N\ge 4$ systems may therefore be seen to be further consequences 
of the {\it matreoshka phenomenon} \c{matr} of \ssd systems; and the 
{\it \sd matreoshka} can even be taken to have infinitely many layers.

The spectra of the $N$-extended \ssd systems consist of the \ym vector
potential $A_{\a\db}$ having \sd field-strength $f_{\a\b}$, a $(\half,0)$
spinor $\l_{i\a}$, and spin ${n\over 2}$ fields 
$\{ \phi_{i_{n+2}\dots i_1 \da_1\dots \da_{n}}; $ $ 0\le n\le (N-2)\}$
transforming according the totally symmetric $(0,{n\over 2})$ \rep s of 
the rotation group  and according to skew-symmetric representations of 
the internal $SL(N)$ automorphism group of the N-extended supersymmetry
algebra.
The \N6 theory, for instance, has the following spectrum transforming
according to an irreducible \rep\ of the \N6 \sp\
$$  A_{\a\db}  \qquad \l_{i\a}  \qquad   W_{ij}
\qquad \x_{ijk\da} \qquad  g_{ijkl\da\db}\qquad\psi_{ijklm\da\db\dg}
\qquad C_{ijklmn\da\db\dg\dd}$$
where $i,j = 1,...,N $ are internal $sl(N)$ indices which we always write
as subscripts, so for instance, the spin 2 field $C_{ijklmn \da\db\dg\dd}$
above is an $sl(6)$ singlet and the spin ${3\over 2}$ field
$\psi_{ijklm \da\db\dg}$ is an $sl(6)$ vector, and these may be more
conveniently denoted if \N6 thus:
$C_{\da\db\dg\dd} = {1\over 6!} \e^{ijklmn} C_{ijklmn \da\db\dg\dd} \quad ,
\quad \psi^i_{\da\db\dg} = {1\over 5!} \e^{ijklmn} \psi_{ijklm \da\db\dg}$.
However, our notation has the advantage of being $N$--independent.
All these fields take values in the Lie algebra of the gauge group and are
linear in the \ym coupling constant, which we absorb into 
the definition of these fields. There is no other coupling constant. 
This means that unlike conventional field theories, where bosonic
fields have dimension $-1$ and fermionic ones dimension $-{3\over 2}$,
the fields in these \sd \ym multiplets have dimensions which decrease
with spin. So although the fields $  A_{\a\db}, \l_{i\a}, W_{ij},
\x_{ijk\da}$ have conventional dimensionalities, the further fields 
$\phi_{i_{n+2}\dots i_1 \da_1\dots \da_{n}}$ of spin ${n\over 2}$ have
dimension $-{(n+2)\over 2}$. This, together with the fact that there
is no coupling constant apart from the (dimensionless) \ym one, renders 
it impossible to write dimensionless action functionals for the $N > 4$ 
theories. The vector potential transforms in the usual inhomogeneous 
fashion, whereas all other fields transform covariantly under gauge 
transformations,
\be\arr
  \d A_{\a\db} &=& - \p_{\a\db}\tau(x) - [A_{\a\db},\tau(x) ]\  ,\\[5pt]
     \d \l_{i\a} &=& [\tau(x) , \l_{i\a}]\  ,\\[5pt]
     & \vdots & \\[5pt]
     \d \fn &=&   [\tau(x) , \fn ]\  .\la{gt}\ea\ee
These are the only gauge--transformations of these fields; there are no
higher--spin gauge--invariances. The latter not being required since 
all the fields apart from the vector potential transform according to
irreducible representations of the rotation group and therefore have no
redundant degrees of freedom. We recall that inhomogeneous gauge 
transformation of a field can be understood as the condition for the 
field to describe a degree of freedom of unique spin \c{op3}.
It is in fact precisely these features of having only one coupling 
constant and one type of gauge invariance which render traditional 
theorems forbidding higher--spin couplings inapplicable to our systems.

The arbitrary $N$ supersymmetry transformations take the form
\be\begin{array}{lll}  \d A_{\a\db}&=& - \eb^i_\db  \l_{i\a}  \\[5pt]
   \d \l_{j\a} & = &  \eta_j^\b  f_{\a\b}
                       + 2 \eb^{i\db}  \n_{\a\db} W_{ij}      \\[5pt]
   \d W_{jk}   & = &  \eta_{[j}^\a \l_{k]\a}
                       + \eb^{i\db} \x_{ijk\db}     \\[5pt]
 \d \x_{jkl\da} & = &  \eta_{[j}^\a \n_{\a\da} W_{kl]}
    + \eb^{i\db} \left(g_{ijkl\da\db} + \e_{\da\db} [W_{i[j},W_{kl]} ] 
 \right) \\[5pt]
 \d g_{jklm\da\db} & = &  \eta_{[j}^\a \n_{\a(\da} \x_{klm]\db)}    \\[5pt]
  && + \eb^{i\dg}\left(\psi_{ijklm\da\db\dg}
  +\e_{\dg(\da} \left( {2\over 3}[W_{i[j},\x_{klm]\db)}]
                  - {1\over 3}[W_{[jk},\x_{lm]i\db)}] \right) \right)
     \\[5pt]  & \vdots & \\[5pt]
 \d \fn &=&  \eta_{[i_n}^\a \n_{\a(\da_n} 
 \phi_{i_{n+1}\dots i_1 \da_1\dots \da_{n-1})} \\[5pt]
  &&  + \eb^{i_{n+3}\da_{n+1}} \left( \gn + \e_{\da_{n+1}(\da_n}
\Z_{i_{n+3}\dots i_1 \da_1\dots \da_{n-1})}  \right)  ,\la{susy}\ea\ee
where  $\Z_{i_{n+3}\dots i_1 \da_1\dots \da_{n-1}}$ is a functional
of fields of spin less than ${n\over 2}$.
\goodbreak
\vskip 10pt
{\bf 1.4 } The \N4 \sd theory was considered by Siegel \c{s}.
The appearance of the additional spin 1 field $g_{ijkl\da\db}$, which
for \N4 is equal to $\e_{ijkl}g_{\da\db}$,
is particularly noteworthy.
This possibilty of nontrivially coupling two mutually independent spin 1
fields is a peculiarity of \sd theories. There is in fact no analogue in
standard (non-\sd) gauge theories, for which the conserved vector current
which acts as a source for the \ym field,
provides all consistent spin 1 couplings (including self-couplings) \c{op2}.
In Minkowski space the two helicities of the gauge field are complex
conjugates of each other. However, for \sd theories (in spaces of
signature (4,0) or (2,2)), the \ym current acts as a source for only
the \sd $(1,0)$ half of the gauge field. This leaves room for a
vector current which acts a source for a $(0,1)$ field;
and the \N4 theory opens the way for such an additional spin--one field.
The source vector current \r{n4} for this new spin--one field is actually
the Noether current 
\be   j_{\a\db ijkl} = -[A_\a^{\;\da} , g_{ijkl\da\db }] +
      \{ \l_{[i\a} , \x_{jkl]\db} \}
                       -  [ W_{[ij}, \n_{\a\db} W_{kl]} ]  \la{cc4}\ee
corresponding to global gauge invariance of the action functional for
the \N4 theory \c{s}
\be S = \int d^4x\  \Tr \e^{ijkl} \left( f^{\da\db}g_{ijkl\da\db}
               + \x_{[ijk}^\da \n_{\a\da} \l^\a_{l]}
	       +  W_{[ij}\square W_{kl]}
	       -  W_{[ij} \{ \l^\a_{k} , \l_{l]\a} \}
	       \right)  .\la{s}\ee
Repeated supersymmetry transformations of this spin-one current yield source
currents for succesively higher-spin fields, which make up successively
higher-$N$ supermultiplets.

The action \r{s} is a component version of the light-cone chiral superspace
action \c{s}  based on the  Lagrangian due to \c{l}
$$ L = \half V^{--} \square V^{--} - 
 {1\over 3}  V^{--} [\p^{\a+}V^{--},\p^+_{\a} V^{--}]  ,$$
which is an N-independent Lagrangian for the arbitrarily extended 
\ssdy equations (see also \c{cd}).
An alternative harmonic superspace action has been presented in \c{e}.

The existence of the invariant
action functional \r{s} for the \N4 theory also gives rise to a conserved
gauge--invariant stress tensor for this theory, yielding a possible source
term for Einstein's equations. This is a conserved tensor for all 
$N\ge 4$ theories, which therefore allow a non-trivial coupling to
gravity, whereas for $N\le 3$ the only possible gravitational source term
is the standard  \ym stress tensor
$$ T_{\alpha{\dot \alpha},\beta{\dot \beta}}
= \Tr f_{{\dot \alpha}{\dot \beta}} f_{\alpha\beta},$$
which vanishes identically for \sd theories. 

\vskip 10pt
{\bf 1.5 }
Conventionally, manifestly supersymmetric forms of extended 
supersymmetric systems take the form of \ss supercurvature 
constraints. In fact, the standard natural set of supercurvature 
constraints describes our arbitrarily--extended systems as well.

In $N$-extended superspace with coordinates 
$ \{ x^{\a\da}, \bt^{i\da}, \t^\a_i \}$, where
$\{ \bt^{i\da}, \t^\a_i ; i=1,\dots N \} $ are odd coordinates and 
$x^{\a\da}$ are standard coordinates on \M, on which the 
{\it component fields} above depend. A {\it \sd superconnection} is 
subject to the following constraints (e.g. \c{ssd})
\bealph{20}
 & \{\hn^i_\a , \hn^j_\b \} &=\  \  0 \\[5pt]
 & [ \hn^i_\a , \hn_{\b \db}  ]  &=\  \   0 \\[5pt]
 & \{\hn^i_\a ,   \hn_{j \db} \} &=\  \   2 \d^i_j \hn_{\a \db} \\[5pt]
 & \{\hn_{i (\da},  \hn_{j \db)} \} &=\  \   0 \\[5pt]
 & [\hn_{i (\da}, \hn_{\b \db)}  ]  &=\  \   0 \\[5pt]
 & [\hn_{\a(\da}, \hn_{\b \db)}  ]  &=\  \   0   
\eealph{20}
The first three conditions allow the choice of a {\it chiral basis}
in which the covariant derivatives take the form
$$\arr 
 &\hn^i_\a &=\  \   \der{\t_i^\a} \\[5pt] 
 &\hn_{i \da} &=\  \   \n_{i \da} + 2 \t_i^\a \n_{\a \da} \\[5pt]
 &\hn_{\a \da}  &=\  \   \n_{\a \da},\ea$$
where $(\n_{i \da}, \n_{\a \da})$ are covariant derivatives in the
{\it chiral subspace} independent of the $\t_i^\a$ coordinates. In this 
basis the single constraint (20d), equivalently written in the form
$$ \{\hn_{i \da},  \hn_{j \db} \} =  \e_{\da\db} \widehat f_{ij}\  ,$$
encapsulates the content of all the other constraints and the odd 
component of the superconnection 
$$ \widehat A_{i \da}(x,\bt,\t) =
  A_{i \da}(x,\bt) + 2 \t_i^\a A_{\a \da}(x,\bt)\   ,$$ 
describes the entire \sd supermultiplet in the form of the curvature
component $ \widehat f_{ij}$, which has a quadratic $\t$-expansion in terms 
of chiral superfields of the form
\be \widehat f_{ij}(x,\bt,\t) = 
 f_{ij}(x,\bt)  + 2 \t^\a_{[i}\f_{j]\a}(x,\bt)  
  + 4\t^\a_i\t^\b_j f_{\a\b}(x,\bt)\  .\la{hatf}\ee
As we shall see, the $\bt$-expansion of $f_{ij}$ yields all the higher
spin fields $\x_{ijk\da}, g_{ijkl\da\db},$ $\psi_{ijklm\da\db\dg},$
$C_{ijklmn\da\db\dg\dd}, \dots $ etc.
 
\vskip 10pt
{\bf 1.6 }
The plan of this paper is as follows. In section 2, we shall show that 
the \sd supercurvature constraints in chiral \ss yield a spectrum of
chiral superfields having the \lc\ fields described above, as well as 
their dynamical equations.
This formulation of \ssdy in chiral \ss is the starting point for the
establishment of the supertwistor correspondence for these systems.
We have previously described \c{matr,pl} a \hs formulation of this
correspondence, which, being $N$--independent, holds in the arbitrary
$N$ case too. This supertwistor correspondence yields a complete
characterisation of the solution space only if the relation between the \ym
superconnection satisfying the supercurvature constraints and the set of
component fields satisfying the component \ssdy \eq s is one-to-one.
Machinery for establishing such equivalences was developed
in \c{hhls}, where the conventional \N3 superconnection constraints were
proven to be completely equivalent to the full \sym \eoms. The method
was later applied to the ten dimensional case too \c{hs}. This method,
which we apply to the \ssdy conditions in sections 3 and 4, also yields a 
very effective mechanism for extracting component information from \sf\  
data. The new \N4 stress tensor is actually a member of a supermultiplet 
of conserved tensors. In fact $N>4$ generalisations of these conserved 
tensors also exist and are presented in section 5.

\goodbreak\vskip 10pt
\section{ Self-duality constraints for the superconnection and the \sf\
\eoms\  }
\vskip 10pt
The $N$-extended \ssdy equations in four dimensional space are most
economically written in $N$-extended {\it chiral superspace}, \sM, with
coordinates $ \{ x^{\a\da}, \bt^{i\da} \}$, where
$\{ \bt^{i\da}; i=1,\dots N \} $ are odd coordinates and $x^{\a\da}$ are
standard coordinates on which the {\it component fields} on \M\ 
depend. For generality, we shall work in the complexified 
setting. Reality conditions appropriate to a (4,0) or (2,2) signature may 
always be imposed. We shall take the extension $N$ to be arbitrary.
Gauge-covariant derivatives in chiral superspace \sM\ take the form
$$\arr         \n_{i \da} &=& \p_{i \da} + A_{i \da}  \\[5pt]
               \n_{\a \da} &=& \p_{\a \da} + A_{\a\da}\  ,\ea$$
where the partial derivatives $\p_{i \da} \equiv \der{\bt^{i \da}},\quad
\p_{\a \da} \equiv \der{x^{\a \da}}$
provide a holonomic basis for \sM; chiral superspace being
torsion-free. The components of the superconnection $( A_{i\da}, A_{\a\da})$
take values in the Lie algebra of the gauge group, their transformations
being parametrised by Lie algebra-valued sections on \sM\  (c.f. \r{gt}):
\bealph{22}
 \d A_{i\da} &=& - \p_{i\da}\tau(x,\bt) - [A_{i\da},\tau(x,\bt) ]
 \la{sgt}\\[5pt]
 \d A_{\a\db} &=& - \p_{\a\db}\tau(x,\bt) - [A_{\a\db},\tau(x,\bt) ]\ .
\eealph{22}
On \sM, the \ssdy conditions take the form of the following
supercurvature constraints
\bealph{23} \{\n_{i (\da},  \n_{j \db)} \} &=&  0 \la{css}\\[5pt]
 [\n_{i (\da}, \n_{\b \db)}  ]  &=&  0  \la{csv}\\[5pt]
 [\n_{\a(\da}, \n_{\b \db)}  ]  &=&  0\  ,\la{cvv}\eealph{23}
or equivalently
\bealph{24}
     \{\n_{i \da},  \n_{j \db}\} &=& \e_{\da\db} f_{ij} \\[5pt]
     [\n_{i  \da}, \n_{\b \db}] &=&  \e_{\da\db} f_{i \b}  \\[5pt]
     [\n_{\a \da}, \n_{\b \db}] &=&  \e_{\da\db} f_{\a\b}\  ,\la{cvv1}
\eealph{24}
where $f_{ij}=f_{ij}(x,\bt)$ is skew--symmetric and 
$f_{\a\b}=f_{\a\b}(x,\bt)$ is symmetric and has the corresponding \M\ 
\ym field-strength $f_{\a\b}(x)$ as its leading component in a
$\bt$--expansion. Henceforth all fields are superfields depending on both
$x^{\a\da}$ and $\bt^{i\da}$ and we distinguish \sf s from their leading
components in a $\bt$-expansion (i.e. ordinary fields on \M) by placing a
circle over the latter. We shall
henceforth thus write all fields of section 1.3, which are leading
components of corresponding \sf s; and we shall denote the latter by the
same letter as the former. Thus $\l_{i\a}$, for instance, will be used to
denote the \sf\ containing $\0\l_{i\a}$ as its $\bt$-independent part.

The \sf\ curvatures (24) are not independent; they are related by
super-Jacobi identities. Firstly, the dimension $-3$ \ji\ implies, in virtue
of the constraint \r{cvv},  the \ym equation
\be  \n^\a_{\;\db} f_{\a\b} = 0  .\la{ym}\ee
for the {\it \sf\ } $f_{\a\b}$. Next, in virtue of the constraints \r{csv}
and \r{cvv}, the dimension $-2\half$ Jacobi identity yields the relationship
\be \n_{i\da} f_{\a\b}    =  \half \n_{(\a\da} f_{i\b)}  .\ee
Multiplying both sides by $\e^{\a\b}$ yields a dynamical equation for
the dimension $-{3\over 2}$ curvature, which allows its identification
with a spinor superfield
$$ \l_{i\a} := f_{i\a}\  ,$$
having \eom\
\be \n^\a_{\;\da} \l_{i\a} =   0\   .\la{el}\ee
Now the \dim{-2} \ji\ says that
\be \n_{\a\da} f_{ij}  =  \n_{i\da} \l_{j\a}\      .\la{bw1}\ee
Defining a scalar \sf\   \be  W_{ij} :=  \half  f_{ij}\   ,\la{dw}\ee
where the rescaling merely serves to bring our notation into 
correspondence with that in the literature, we obtain the \eom\
\be \square W_{ij}  = \half  \{ \l^\a_i, \l_{j\a} \}\  ,\la{ew}\ee 
where the covariant d'Alembertian is defined by
$\square = \half\n^{\a\db}\n_{\a\db}$. 
The curvature constraints (23), combined with the definitions
(24), may therefore also be written
\be\arr
\{\n_{i \da}, \n_{j \db}\} =& 2\e_{\da \db} W_{i j} \\[8pt]
[\n_{i \da}, \n_{\b \db}] =& \e_{\da \db}   \l_{i \b}   \\[8pt]
[\n_{\a \da}, \n_{\b \db}] =& \e_{\da \db}  f_{\a \b}\  .
\la{con}\ea\ee

Now  the \dim{-{3\over 2}} \ji\  tells us that the spinorial derivative of the
scalar \sf, $\n_{i\da} W_{jk}$, is a \sf , totally skew-symmetric in $ijk$:
\be \n_{i\da} W_{jk}    =  \x_{ijk \da}\           .\la{bx}\ee
Acting on both sides by $\n_\a^{\;\da}$ yields
$$\arr  \n_\a^{\;\da} \x_{ijk \da} &=& \n_\a^{\;\da} \n_{i\da} W_{jk} \\[5pt]
      &=& 2 [\l_{i\a},W_{jk} ] + \half \n_{i\da} \n_j^{\da}\l_{k\a}\quad
      \mbox{using (24b) and \r{bw1}.}\ea$$
Now the second term on the right is equal to
$$\arr  && - 2 [ W_{ij} , \l_{k\a} ] - \half \n_j^\da \n_{i\da}\l_{k\a} \\[5pt]
 &\quad& = 2 [ \l_{k\a} , W_{ij} ] -  \n_j^\da \n_{\a\da} W_{ik}
      \quad\mbox{using \r{bw1},}\\[5pt]
&\quad& = 2 [ \l_{k\a} , W_{ij} ] -  2 [ \l_{j\a} , W_{ik} ]
        - \n_\a^{\;\da} \x_{ijk \da}
      \quad\mbox{using \r{con},\r{bx}.}\ea $$
We therefore have the \eom\ for the \sf\ \footnote{All our
(skew-)symmetrisations are with weight one. For instance,
$ [ \l_{[i\a} , W_{jk]} ] \equiv [ \l_{i\a} , W_{jk}] +
[ \l_{j\a} , W_{ki}] + [ \l_{k\a} , W_{ij}] .$}$x_{ijk\da}$,
\be  \n_\a^{\;\da} \x_{ijk\da } =  [ \l_{[i\a} , W_{jk]} ]  .\la{ex}\ee
Action of the covariant derivative $\n^\a_\db$ yields the wave equation
\be  \square \x_{ijk\da } =  [ \l_{[i}^\a , \n_{\a\da} W_{jk]} ]  ,\la{wx}\ee
since $ \n^\g_\db \n_\g^{\;\da} = - \d^\da_\db \square $ in virtue of the 
curvature constraint \r{cvv}. 
Now using \r{bx} we have 
\be 3 \n_{i\da} \x_{jkl\db} = \n_{i\da} (\n_{[j\db} W_{kl]})  
=  2 \e_{\da\db} [ W_{i[j}, W_{kl]}] - \n_{[j\db} \x_{kl]i\da}\ .\la{bbg}\ee
So
\be 4 \n_{i\da} \x_{jkl\db} = - \n_{[j\db} \x_{kl]i\da} + 
         \n_{i\da} \x_{jkl\db} +  2 \e_{\da\db} [ W_{i[j}, W_{kl]}]\  .\ee
Symmetrising in $\da , \db$ yields the definition of a rank-$4$  
spin-one \sf\  $g_{ijkl\da\db}$,
\be  4 \n_{i(\da} \x_{jkl\db)} =\  - \n_{[j(\da} \x_{kli]\db)} 
                              =\ 8 g_{ijkl\da\db}\  .\la{dg}\ee
Eq. \r{bbg} also imples that 
$$ 2 \n_{i\da} \x_{jkl\db} = - \n_{[j\db} \x_{kl]i\da} - 
            \n_{i\da} \x_{jkl\db} +  2 \e_{\da\db} [ W_{i[j}, W_{kl]}]\  $$
and tracing over the spinor indices yields
$$ 2  \n^\db_i \x_{jkl\db} 
      = \n^\db_{[j} \x_{kli]\db} + 4 [ W_{i[j}, W_{kl]}]\  .$$
This implies that $\n^\db_{[j} \x_{kli]\db}=0$, identically, and we
therefore have the \ss relationship between $g_{ijkl\da\db}$ and 
lower-spin fields,
\be\arr \n_{i\da} \x_{jkl\db} &\equiv & \half \n_{i(\da} \x_{jkl\db)} +
                        \half \e_{\da\db}  \n^\dg_i \x_{jkl\dg} \\[5pt]
 &=& g_{ijkl\da\db} + \e_{\da\db} \Z_{ijkl}\   
\hbox{, where }\quad \Z_{ijkl} = [ W_{i[j}, W_{kl]}] \la{bg}\ea\ee
Now, the action of $\n_\a^{\;\db}$ on both sides and use of 
\r{bw1} and \r{con}-\r{ex} yields
$$\arr
\n_\a^{\;\db} g_{ijkl\da\db }  &=&  
\{ [ \n_\a^{\;\db} , \n_{i\da} ], \x_{jkl\db} \} 
  + \{ \n_{i\da} , \n_\a^{\;\db} \x_{jkl\db} \} 
  - \n_{\a\da}[ W_{i[j}, W_{kl]}] \\[5pt]
&=& \{ \l_{i\a}, \x_{jkl\da} \} + \{ \n_{i\da},[\l_{[j\a} , W_{kl]}]
 - [ \n_{\a\da} W_{i[j}, W_{kl]}] - [ W_{i[j},  \n_{\a\da} W_{kl]}]\\[5pt]
&=& \{ \l_{i\a}, \x_{jkl\da} \} - \{ \l_{[j\a}, \x_{kl]n\da} \} +
 [ \n_{\a\da} W_{i[j}, W_{kl]}] - [ W_{i[j},  \n_{\a\da} W_{kl]}]\\[5pt]
&=& \{ \l_{[i\a}, \x_{jkl]\da} \} + [ \n_{\a\da} W_{[ij}, W_{kl]}]\  
  .\ea$$ 
So the \eom\ for the spin-one \sf\  $g_{ijkl\da\db}$ follows;
\be   \n_\a^{\;\db} g_{ijkl\da\db }  =  J_{ijkl\a\da}\   ,\la{eg}\ee
where the current \be   J_{ijkl\a\da} =  \{ \l_{[i\a} , \x_{jkl]\da} \}
                       -  [ W_{[ij}, \n_{\a\da} W_{kl]} ]  .\la{c4}\ee
is covariantly conserved,
\be
\n^{\a\db} J_{ijkl \a\db} = 0\    .\la{s1cc}\ee
{\it Proof.} In virtue of \r{el},
$$  \n^{\a\db} J_{ijkl \a\db} = \{ \l_{[i\a} , \n^{\a\db} \x_{jkl]\db} \}
     -  [ \n^{\a\db} W_{[ij}, \n_{\a\db} W_{kl]} ] 
     -  2 [ W_{[ij}, \square W_{kl]} ] $$
The second term is identically zero and using the \eoms\ \r{ew} and \r{ex}
we see that the rest of the right-hand side also vanishes
in virtue of the \jis. \square

Now, action of the covariant derivative $\n^\a_\dg$ on \r{eg} yields
$$  \n^\a_\dg  \n_\a^{\;\db} g_{ijkl\da\db } =  \n^\a_\dg J_{ijkl\a\da}
 = \half \e_{\dg\da} \n^{\a\db} J_{ijkl \a\db} 
  + \half \n^\a_{(\dg} J_{ijkl\a\da)} = \half \n^\a_{(\dg} J_{ijkl\a\da)}\ 
 ,$$ 
using \r{s1cc}. We therefore obtain the wave equation
\be \square g_{ijkl\da\db } =
     -\half \{ \l_{[i\a} , \n^\a_{(\da} \x_{jkl]\db)} \}
          + \half  [\n^\a_{(\da} W_{[ij}, \n_{\a\db)} W_{kl]} ]  .\la{wg}\ee
Remarkably, the equations of motion obtained hitherto for the partial
supermultiplet \quad 
$\{ A_{\a\db},$ $\l_{i\a},$  $W_{ij}, \x_{ijk\da}, g_{ijkl\da\db} \} $ \quad
are Euler-Lagrange equations for the simultaneous variation of the
${\scriptscriptstyle \pmatrix{ N\cr 4}}$ \sf\  functionals
\be \L_{ijkl} =  \Tr  \left( f^{\da\db}g_{ijkl\da\db}
               + \x_{[ijk}^\da \n_{\a\da} \l^\a_{l]}
	       +  W_{[ij}\square W_{kl]}
	       -  W_{[ij} \{ \l^\a_{k} , \l_{l]\a} \}
	       \right)  ,\la{l4}\ee
whose leading ($\t$-independent) terms 
yield, for \N4, the action functional
$ S=\int dx^4 \e^{ijkl} \L_{ijkl} $, i.e. \r{s}.

Consider now, \be\arr
  \n_{i\da} g_{jklm\db\dg}  &\equiv& 
                 {1\over 3} \left(\n_{i(\da} g_{jklm\db\dg)}
                 + \e_{\da(\db} \n_i^\dd g_{jklm\dg)\dd}\right) 
\\[5pt]		 &=& \psi_{ijklm\da\db\dg}
		 + \e_{\da(\db} \Z_{ijklm\dg)}\   ,\la{bh1}\ea\ee
which defines a spin ${3\over 2}$ \sf\   $\psi_{ijklm\da\db\dg}$. Now
from \r{dg} and using \r{bx}, \r{bg},
$$\arr 2 \n_i^\dg g_{jklm\db\dg} &=&  
       \n_i^\dg (\n_{j(\db} \x_{klm\dg)}) \\[5pt] 
&=& [[\n_i^\dg , \n_{j(\db}], \x_{klm\dg)} ] - 
\n_{j\db} (\n_i^\dg  \x_{klm\dg})  - \n_{j\dg} (\n_i^\dg  \x_{klm\db}) 
\\[5pt] &=&  6 [ W_{ij} , \x_{klm\db} ] 
  - 2 \n_{j\db} ( [ W_{i[k} ,W_{lm]}])
   + \n_j^\dg (g_{iklm\dg\db}  + \e_{\dg\db} [ W_{i[k}, W_{lm]}]) 
\\[5pt] &=&  6 [ W_{ij} , \x_{klm\db} ] 
            - 3 [\x_{ji[k\db},W_{lm]} ] - 3 [ W_{i[k},\x_{lm]j\db}] 
               + \n_j^\dg g_{iklm\db\dg}\     .\ea$$ 
So on adding parts symmetric and skew-symmetric in $i,j$ we obtain
\be  \n_i^\dg g_{jklm\db\dg}
 = 2 [ W_{i[j} , \x_{klm]\db}]  - [ W_{[jk} , \x_{lm]i\db} ]\  ,\la{sg}\ee
yielding, from \r{bh1}, the relation between  $\psi_{ijklm\da\db\dg}$ 
and lower-spin fields,
\be \n_{i\da} g_{jklm\db\dg} =  \psi_{ijklm\da\db\dg} +
	    \e_{\da(\db}\left( {2\over 3} [ W_{i[j} , \x_{klm]\dg)} ]
	  - {1\over 3} [ W_{[jk} , \x_{lm]i\dg)} ]  \right)   .\la{bh}\ee
The spin ${3\over 2}$ \eom\ follows in virtue of lower-spin equations on
application of $\n_\a^{\;\dg}$ to both sides, namely,
\be \n_\a^{\;\dg} \psi_{ijklm\da\db\dg } =  J_{ijklm\a\da\db}\  ,\la{eh}\ee
with \be J_{ijklm\a\da\db} =
  [ \l_{[i\a} , g_{jklm]\da\db} ]
        + {2\over 3}[\n_{\a(\da} W_{[ij},  \x_{klm]\db)} ]
	- {1\over 3}[ W_{[ij}, \n_{\a(\da} \x_{klm]\db)} ]  ,\la{c5}\ee
a covariantly conserved current satisfying
\be   \n^{\a\da} J_{ijklm\a\da\db} = 0  \ee
in virtue of lower spin \eoms. Similarly,
\be \n_{i\da} \psi_{jklmn\db\dg\dd} =   C_{ijklmn\da\db\dg\dd} +
	    \e_{\da(\db} \Z_{ijklmn\dg\dd)}\   \la{bc}\ee
where the spin 2 \sf\ is defined by
\be C_{ijklmn\da\db\dg\dd}
         = {1\over 4}\n_{i(\da} \psi_{jklmn\db\dg\dd)}
	   \la{dc}\ee
and \be\arr
	\Z_{ijklmn\db\dg} &\equiv & {1\over 4} \n_i^\da \psi_{jklmn\da\db\dg}
      = {1\over 12} \n_i^\da \left( \n_{j(\da} g_{klmn\db\dg)}\right)
\hbox{\quad from \r{bh1},}  \\[5pt]
      &=&  {1\over 6} [ \x_{i[jk\db} , \x_{lmn]\dg} ]  +
            {1\over 2} [ W_{i[j} , g_{klmn]\db\dg}] +
            {1\over 6} [ W_{[jk} , g_{lmn]i\db\dg} ]\   \ea\ee
using \r{sg}. Covariant differentiation of both sides of \r{bc} yields the 
spin 2 dynamical \eq\  
\be  \n_\a^{\;\dd} C_{ijklmn\da\db\dg\dd} = J_{ijklmn\da\db\dg}\   ,\ee
with the covariantly conserved current 
\be\arr    J_{ijklmn\a\db\dg\dd}
   & = & \{ \l_{[i\a} , \psi_{jklmn]\db\dg\dd} \}
	+ {1\over 6}\{ \x_{[ijk(\db}, \n_{\a\dg} \x_{lmn]\dd)} \} \\[6pt]
       & &\quad + \half [\n_{\a(\db} W_{[ij},  g_{klmn]\dg\dd)} ]
    - {1\over 6}[ W_{[ij}, \n_{\a(\db} g_{klmn]\dg\dd)} ]\  .\la{c6}\ea\ee

By iteration of this procedure,  \sf s
$\phi_{i_n\dots i_1 \da_{n-2}\da_1\dots \da_{n-1} }$
of $sl(N)$ rank $n$ of spin $s= {(n-2)\over 2}$ may be produced for higher
spins. The super-\jis\ and the constraints \r{con} admit such \sf s for 
every $n\le N$ and for arbitrarily large $N$.
We therefore have the following \ss recursion relations between $sl(N)$ 
rank $n$ and rank $(n+1)$ \sf s:
\bealph{55}
 \n_{i\da} f_{\a\b}    &=&  \half \n_{(\a\da} \l_{i\b)}    \la{bl}\\[5pt]
  \n_{i\da} \l_{j\a}   &=& 2 \n_{\a\da} W_{ij}             \la{bw}\\[5pt]
\n_{i\da} W_{jk}    &=&  \x_{ijk \da}\                            \\[5pt] 
 \n_{i_{n+2}\da_1} \phi_{i_{n+1}\dots i_1 \da_2\dots \da_{n} }
	     &=&  \phi_{i_{n+2}\dots i_1 \da_1\dots \da_{n} }
 + \e_{\da_1(\da_2} \Z_{i_{n+2}\dots i_1\da_3\dots \da_{n})}\
  ,\quad n\ge 2 ,\la{bf}\eealph{55}
where the spin ${n\over 2}$ \sf\ is defined recursively by
\be \phi_{i_{n+2}\dots i_1 \da_1\dots \da_{n} } = {1\over n}
\n_{i_{n+2}(\da_1} \phi_{i_{n+1}\dots i_1 \da_2\dots \da_{n} )}\
 \la{df}\ee and
\be \Z_{i_{n+2}\dots i_1\da_1\dots \da_{n-2}}  = {1\over n}
\n_{i_{n+2}}^{\da_{n-1}}\phi_{i_{n+1}\dots i_1\da_1\dots \da_{n-1}}\  \ee
are completely determined in terms of lower spin superfields.
The first two \eq s in this series are given by \r{bg} and \r{bh} with 
$$\arr \Z_{ijkl} &=& [ W_{i[j}, W_{kl]}]\  \\[5pt]
\Z_{ijklm\dg} &=& {2\over 3} [ W_{i[j} , \x_{klm]\dg} ]
	  - {1\over 3} [ W_{[jk} , \x_{lm]i\dg} ]  .\ea$$
In the general case, the action of $\n_{i_{n+3}}^{\da_n}$ 
on \r{df} yields
\be\arr && n \n_{i_{n+3}}^{\da_n} \phi_{i_{n+2}\dots i_1\da_1\dots \da_{n}} = 
2(n+1)\  [ W_{i_{n+3}i_{n+2}} , \phi_{i_{n+1}\dots i_1\da_1\dots \da_{n-1}}]
  \\[5pt]
 && \quad + 
\n_{i_{n+2}}^{\da_n} \phi_{i_{n+3}i_{n+1}\dots i_1\da_1\dots \da_{n}}
  - (n+1)\ 
  \n_{i_{n+2}(\da_1} \Z_{i_{n+3}i_{n+1}\dots i_1 \da_2\dots \da_{n-1})}\
   .\ea\ee
On adding the parts symmetric and antisymmetric in $i_{n+2}i_{n+3}$ we
obtain
\be\arr
&&  (n+1)\  \Z_{i_{n+3}\dots i_1 \da_1\dots \da_{n-1}}\  =\ 
  \n_{i_{n+3}}^{\da_n} \phi_{i_{n+2} \dots i_1 \da_1\dots  \da_{n}} 
\\[5pt] && = 2\ 
[ [ W_{i_{n+3}i_{n+2}} , \phi_{i_{n+1}\dots i_1 \da_1 \dots \da_{n-1}}]
   -\half
  \n_{[i_{n+2}(\da_1} \Z_{i_{n+3}]i_{n+1}\dots i_1 \da_2\dots \da_{n-1})}\
  \\[5pt] && \quad -{n+1\over 2n-2}\ 
 \n_{(i_{n+3}(\da_1} \Z_{i_{n+2})\dots i_1 \da_2\dots \da_{n-1})}\  .\ea\ee

The relations (55) contain both dynamical as well as kinematical information.
As we have explicitly seen for the lower spin fields, the dynamical content 
may be extracted by covariant differentiation of (55) and use of the 
constraints \r{con}. In general, the higher-spin \sf s  have \eoms\
\be \n_\a^{\da_n} \phi_{i_{n+2}\dots i_1 \da_1\dots \da_{n} }  = 
            J_{i_{n+2}\dots i_1 \a\da_1\dots \da_{n-1} }\   ,\la{e}\ee
where 
\be\arr    J_{i_{n+2}\dots i_1 \a\da_1\dots \da_{n-1} } &=&
\n_{i_{n+2}\da_1} 
(\n_\a^{\da_n} \phi_{i_{n+1}\dots i_1 \da_2\dots \da_{n}}) \\[5pt]
&& + [[ \n_\a^{\da_n} , \n_{i_{n+2}\da_1} ],
 \phi_{i_{n+1}\dots i_1 \da_2\dots \da_{n} }] - 
  \n_{\a(\da_1} \Z_{i_{n+2}\dots i_1\da_2\dots \da_{n-1})} \\[8pt]
&=& \n_{i_{n+2}\da_1}  J_{i_{n+1}\dots i_1 \a\da_2 \dots \da_{n-1} }
 + [\l_{i_{n+2}\a}, \phi_{i_{n+1}\dots i_1 \da_1\dots \da_{n-1} }] 
  \\[5pt] 
&&  -  \n_{\a(\da_1} \Z_{i_{n+2}\dots i_1\da_2\dots \da_{n-1})}\   
 ;\la{je}\ea\ee
a recursion relation for the higher spin source currents. The latter
therefore allow explicit construction in an iterative fashion, starting 
from the known ones above and using the relations (55) in order to determine
$\n_{i_{n+2}\da_1}  J_{i_{n+1}\dots i_1 \a\da_2 \dots \da_{n-1} }$.
We therefore see that the constraints (23), or equivalently \r{con}, in
virtue of the super-Jacobi identities, recursively reveal an unending chain
of dynamical equations for \sf s of increasing spin beginning with the spin
$\half$ \eq\ for $\x_{ijk\da}$. These equations moreover have the form
of the interacting \def\df{Dirac--Fierz } \df equations \r{int} with 
source currents,
$J_{i_{n+2}\dots i_1 \g \da_1\dots \da_{n-1}}$, which are functionals
of all fields of spin $\le {n\over 2}$.
Consistency of the linear equations of motion \r{e} 
requires covariant constancy of these currents,
\be \n^{\g\da_{1}} J_{i_{n+2}\dots i_1 \g \da_1\dots \da_{n-1}}
                                   = 0\  ;\la{cc}\ee
conditions satisfied non-trivially in virtue of lower-spin equations 
of motion. The equations  of motion \r{e} have the general form
\be\arr
\p_\a^{\;\da_{n}} \phi_{i_{n+2}\dots i_1\da_1\dots \da_{n}}
  &=& j_{i_{n+2}\dots i_1 \g \da_1\dots \da_{n-1}}  \\[5pt]
  &=& J_{i_{n+2}\dots i_1 \g \da_1\dots \da_{n-1}}
 - [ A_\a^{\;\da_{n}},
         \phi_{i_{n+2}\dots i_1\da_1\dots \da_{n}}]\
    ,\la{cur}\ea\ee
where the currents on the right (which are symmetric in their dotted 
indices) are divergence--free,
\be
\p^{\a \da_{1}} j_{i_{n+2}\dots i_1 \a \da_1\dots \da_{n-1}}
                   =\ 0  .\ee
The spin 1 current in this chain,
the source $j_{ijkl\a\db}$ for the spin 1 field $g_{ijkl\da\db}$  
is precisely the Noether current corresponding
to global gauge invariance of the funtional  $\L_{ijkl}$  \r{l4}.

We have seen that the constraints imply not only the existence of 
higher--spin \sf s but also their \eoms. In fact, the \sf\ \eoms\ 
are not only implied by the supercurvature constraints \r{con}, but
are actually {\it equivalent} to them. Whereas above we have assumed
the constraints in order to derive the \sf\ \eoms, the former may 
instead be seen to arise as consequences of the latter. This 
converse implication follows from the linear equations 
for $\phi$:
\be\arr
\n_{i_{n+3}}^{\da_{n}}\phi_{i_{n+2}\dots i_1\da_1\dots \da_{n}} 
&=&  (n+1)\  \Z_{i_{n+3}\dots i_1\da_1\dots \da_{n-1}} \\[5pt]
 \n_\a^{\da_n} \phi_{i_{n+2}\dots i_1 \da_1\dots \da_{n} }  &=& 
            J_{i_{n+2}\dots i_1 \a\da_1\dots \da_{n-1} }\  .\ea\ee
The consistency conditions for these equations are precisely
the constraint equations (23), or equivalently \r{con}. It is in fact
sufficient to consider the first--order spin-one \sf\ \eq s \r{sg}
and \r{eg}, for which eqs. (23) are the compatibility conditions.

Explicitly, covariantly diffentiating \r{eg} with respect to $\n_\b^{\;\da}$
yields 
$$  [\n_\b^{\;\da} ,\n_\a^{\;\db}] g_{ijkl\da\db }  =
      \n_\b^{\;\da} J_{ijkl\a\da} - \n_\a^{\;\db} J_{ijkl\b\db} = 
      \e_{\b\a} \n^{\g\dg} J_{ijkl\g\dg} =  0    $$
in virtue of \r{s1cc}. Therefore, since $g_{ijkl\da\db}$ is symmetric in
$\da,\db$, \r{cvv} follows as a compatibility condition for \r{eg}.  
Similarly, acting with the spinorial derivative on \r{sg} yields
$$\arr  \n_p^{\;\da}\left(\n_n^{\;\db} g_{ijkl\da\db} \right) &=&
2 \{ \x^\da_{pn[i},\x_{jkl]\da} \} + \{ \x^\da_{n[ij},\x_{kl]p\da} \} 
\\[5pt]&& + 2 [W_{n[i},[W_{|p|j},W_{kl]}]] - [W_{[ij},[W_{|p|k},W_{l]n}]] 
 + [W_{[ij},[W_{kl]},W_{pn}]]\  ,\ea$$
which implies
$$ \{ \n_p^{\;\da} , \n_n^{\;\db} \} g_{ijkl\da\db} = 0 ,$$
a relation equivalent to the curvature constraint \r{css} since
$g_{ijkl\da\db}$ is symmetric in its spinor indices. Finally, 
$$\arr && \n_\a^{\;\da} ( \n_n^{\;\db}  g_{ijkl\da\db})
  - \n_n^{\;\db} (\n_\a^{\;\da} g_{ijkl\da\db}) =
 \n_\a^{\;\da} (2 [ W_{n[i} ,\x_{jkl]\da}] - [ W_{[ij} ,\x_{kl]n\da}]) 
 -  \n_n^{\;\db} J_{ijkl\a\db} \\[5pt] && =
2 [W_{n[i},[\l_{j\a},W_{kl]}]] - 2 [W_{[ij},[\l_{k\a},W_{l]n}]] 
 - 2 [\l_{\a [i},[W_{jk},W_{l]n}]]\  \equiv\ 0 \ea$$
in virtue of the \jis, so the curvature constraint \r{csv} follows.

\goodbreak
\vskip 10pt
\section{ The $\D$-gauge and component expansions of \sf s}
\vskip 10pt

Evaluating the  \sf\ \eoms\ of the previous section at $\bt=0$ yields
\eoms\ of the identical form for the \lc\ fields, with all fields 
including the connection in $\n_{\a\db}$ being ordinary fields on \M\  
and transforming according to $\bt$-independent gauge-transformations 
\r{gt}. Explicitly, we have
\be\arr
 \0f_{\da\db} &=& 0  \\[8pt]
 \0\n^{\a\db} \0\l_{i\a}& = & 0 \\[8pt]
 \0\square \0W_{ij} &=& \half \{ \0\l^\a_i, \0\l_{j\a} \} \\[8pt]
 \0\n_\g^{\;\da} \0\x_{ijk\da } &=&  [ \0\l_{[i\g} , \0W_{jk]} ]  \\[8pt]
 \0\n_\g^{\;\da_{n}} \0\phi_{i_{n+2}\dots i_1\da_1\dots \da_{n}}
  &=& \0J_{i_{n+2}\dots i_1 \g \da_1\dots \da_{n-1}}
  \quad ;\ 4\le (n+2)\le N     .\la{comp}\ea\ee
In the previous section we demonstrated the equivalence of the 
supercurvature constraints (23) to the \sf\ relations (55) and the 
set of \sf\ \eoms\  \r{cvv1}, \r{el}, \r{ew}, \r{ex} and \r{e}.  
In fact we shall prove that:
\vskip 10pt
{\it The following three sets of data are pairwise equivalent, 
up to gauge transformations}.
\vskip 10pt

{\it (i)} The superconnection $\{ A_{i\da}, A_{\a\da} \}$ on \sM\  
subject to the supercurvature constraints (23) (or equivalently \r{con}).

{\it (ii)} Superfields $\{ A_{\a\db}, \l_{i\a}, W_{ij}, \x_{ijk\da},
\phi_{i_{n+2}\dots i_1 \da_1\dots \da_{n}};  2\le n\le (N-2) \}$ on \sM\  
satisfying the \sf\ relations (55), which imply the \sf\ \eoms.

{\it (iii)} The set of component fields $\{ \0A_{\a\db}, \0\l_{i\a},
\0\phi_{i_{n+2}\dots i_1 \da_1\dots \da_{n}};  0\le n\le (N-2) \}$ on \M\  
satisfying the component \ssdy equations \r{comp}.
\vskip 10pt

We have already proven that from a superconnection satisfying (23),
we may recursively construct \sf s $\{ f_{\a\b}, \l_{i\a}, W_{ij}, 
\x_{ijk\da}, \phi_{i_{n+2}\dots i_1 \da_1\dots \da_{n}}; 2\le n\le (N-2) \}$
which automatically satisfy the \sf\ \eoms\ \r{cvv}, \r{el}, \r{ew}, 
\r{ex} and \r{e}. To prove the remaining equivalences, we shall closely 
follow the technique for such
equivalence proofs developed by Harnad et al in \c{hhls}, where the
equivalence between the conventional superspace constraints and the \N3
field equations was given. A similar proof for the d=10 \sym theory was
given in \c{hs}. To obtain $(i)$ or $(ii)$ from $(iii)$, we need to be able
to reconstruct the \sf\ data on \sM\  from the \lc\ fields on \M; and the 
proof of the inverse implications requires \te s of the \sf\ data. Both
reconstruction of \sf s from \lc s, as well as the \te\ of the latter,
clearly require some gauge--fixing, for gauge transformations of \sf s have
parameters depending on the coordinates $\{x,\bt\}$ of \sM, whereas the
component fields have gauge transformations depending only on the
$x$-coordinates of \M. In order to perform \te s or reconstructions of
the \sf s  we clearly need to choose a gauge for them which eliminates all
$\bt$-dependence of their transformation parameters. Following \c{hhls,hs}
we use a `transverse' gauge condition on the odd components of the
superconnection which effectively eliminates the local gauge freedom
associated with the $\bt$-coordinates, viz.
\be \bt^{i\da} A_{i\da} = 0\  .\la{g}\ee
This is tantamount to the requirement that the Euler operator measuring the
degree of homogeneity in the $\bt$ variables is equal to its gauge
covariantisation, i.e.
\be \D \equiv \bt^{i\da}\der{\bt^{i\da}} = \bt^{i\da} \n_{i\da}\  .\la{eu}\ee
Contracting \r{sgt} with $\bt^{i\da}$ we see that in this gauge the parameter
of gauge transformations satisfies $  \D \tau(x,\bt) = 0 ,$
and is therefore homogeneous of degree zero in the odd variables; i.e. it
is $\bt$-independent. This gauge is therefore a suitable one in which to
perform \te s. The condition \r{g} implies no restriction on the
$x$-dependence of the parameters and the \M\ gauge transformations of the
component fields therefore remain intact.

In this gauge, the implication  $(ii)\Rightarrow (iii)$ trivially follows on
evaluating all the \sf s  at $\bt =0$.

Now, in virtue of the fact that the Euler operator $\D$ is gauge--covariant
in this gauge, we can immediately write down its action on all the
\sf s by contracting the \ss relations (55) by $\bt^{i\da}$:
\be\arr
 \D f_{\a\b}    &=&  \half\ \bt^{i\da} \n_{(\a\da} \l_{i\b)}\la{rl}\\[5pt]
 \D \l_{j\a}    &=& 2\ \bt^{i\da}      \n_{\a\da} W_{ij}    \la{rw}\\[5pt]
 \D W_{jk}      &=& \bt^{i\da}        \x_{ijk \da}         \la{rx}\\[5pt]
 \D \x_{jkl\db}
 &=& \bt^{i\da} ( g_{ijkl\da\db}  + \e_{\da\db} [ W_{i[j}, W_{kl]}])
 \la{rg}\\[5pt]
 \D g_{jklm\db\dg} &=& \bt^{i\da} \psi_{ijklm\da\db\dg} -
    \bt^i_{(\db}\left( {2\over 3} [ W_{i[j} , \x_{klm]\dg)} ]
  - {1\over 3} [ W_{[jk} , \x_{lm]i\dg)} ]  \right)   \la{rh}\\[5pt]
  \D h_{jklmn\db\dg\dd} &=& \bt^{i\da} C_{ijklmn\da\db\dg\dd} -
    \bt^i_{(\db} \Z_{ijklmn\dg\dd)}    \\[5pt]
  \D \phi_{i_{n+1}\dots i_1 \da_1\dots \da_{n-1} } &=&
 \bt^{i_{n+2}\da_n} \phi_{i_{n+2}\dots i_1 \da_1\dots \da_{n} } -
 \bt^{i_{n+2}}_{(\da_1} \Z_{i_{n+2}\dots i_1\da_2\dots \da_{n-1})}
 .\la{rec}\ea\ee
Further, contracting the constraints \r{con} by $\bt^{i\da}$, we obtain
\be\arr
(1+\D)A_{i\da} &=&  \  2\ \bt^j_\da  W_{ij}     \\[5pt]
  \D A_{\a\da} &=&  \   -\ \bt^i_\da  \l_{i\a}\  ,\la{reca}\ea\ee
or equivalently 
\be\arr
 [\D , \n_{i\da} ] + \n_{i\da} &=&  2\ \bt^j_\da  W_{ij}\\[5pt]
 [\D , \n_{\a\da}] &=&  -\ \bt^i_\da  \l_{i\a}\  .\la{recd}\ea\ee
Now, since the action of the Euler operator $\D$ on any polynomial in $\bt$
yields the same polynomial with each term multiplied by its degree of
$\bt$-homogeneity, these relations actually determine the \sf s from their
\lc s uniquely. In a \te, the $k$-th order terms on the left-hand sides of
\r{rec} are given by the $(k-1)$-th order terms of the \sf\ expressions
multiplying $\bt$
on the right-hand sides. These relations therefore recursively define the
\sf s from their \lc s; and they do so in a unique manner. In fact the
$\D$-recursions \r{rec} encode the non-dynamical content of the relations 
(55); and similarly \r{reca} contain the non-dynamical part of the
constraints \r{con}. By repeated application of \r{rec} and \r{reca}, 
the leading components may be seen to determine the entire $\bt$-expansion 
of the \sf s. The leading terms are:
\be\arr
f_{\a\b}   &=&  \0f_{\a\b}  + \half \bt^{i\da}   \0\n_{(\a\da} \0\l_{i\b)}
 + \quad \dots \\[8pt]
\l_{j\a}   &=&  \0\l_{j\a}  + 2 \bt^{i\da} \0\n_{\a\da} \0W_{ij}
 + \bt^{i\da}\bt^{l\db}\left(\0\n_{\a\da} \0\x_{lij\dg}
                          - \e_{\da\dg}[ \0\l_{l} , \0W_{ij}] \right)
 + \quad \dots \\[8pt]
W_{jk}     &=&  \0W_{jk}    + \bt^{i\da}   \0\x_{ijk \da}
  + \half \bt^{i\da}\bt^{l\db}\left(\0g_{lijk\da\db}
                        + \e_{\da\db}[\0W_{l[i},\0W_{jk]}]\right)
 + \quad \dots \\[8pt]
\x_{jkl\db}&=&  \0\x_{jkl\db} +
     \bt^{i\da} \left(\0g_{ijkl\da\db}  
               + \e_{\da\db} [ \0W_{i[j}, \0W_{kl]}] \right) 
\\[5pt]&&\quad
+ \half \bt^{i\da} \bt^{m\dg} \0\psi_{mijkl\da\db\dg}
+ \bt^{i\da} \bt^m_\db \left([ \0W_{[jk} , \0\x_{l]mi]\da} ]
                    - {1\over 3}[ \0W_{[mi} , \0\x_{jkl]\da} ]\right)
\\[5pt]&&\quad + \bt^{i\da} \bt^m_\da [ \0W_{i[j} , \0\x_{kl]m\db} ]  
 + \quad \dots \\[8pt]
g_{jklm\db\dg} &=& \0g_{jklm\db\dg} + \bt^{i\da} \0\psi_{ijklm\da\db\dg} -
    \bt^i_{(\db}\left( {2\over 3} [ \0W_{i[j} , \0\x_{klm]\dg)} ]
  - {1\over 3} [ \0W_{[jk} , \0\x_{lm]i\dg)} ]  \right) \\[5pt]&&\quad
+\half \bt^{i\da} \bt^{n\dd} \0C_{nijklm\da\db\dg\dd}
 + \quad \dots ,\la{exp}\ea\ee   
and in general,
$$\arr \fn &=& \0\phi_{i_{n+2}\dots i_1\da_1\dots \da_{n}}
 + \bt^{i_{n+1}\da_{n+1}} \0\phi_{i_{n+3}\dots i_1 \da_1\dots \da_{n+1}}
 \\[5pt]&&\quad -
 \bt^{i_{n+2}}_{(\da_1} \0\Z_{i_{n+2}\dots i_1\da_2\dots \da_{n-1})}
 + \quad \dots \ea$$
and for the superconnection components 
\be\arr
A_{j\da} &=& \bt^{k\da} \0W_{jk} 
 + {2\over 3}\bt^{k\da} \bt^{i\da}\0\x_{ijk\da} 
 \\[5pt] &&\quad + {1\over 4} \bt^{k\da}\bt^{i\da}\bt^{l\db}
         \left(\0g_{lijk\da\db}+\e_{\da\db}[\0W_{l[i},\0W_{jk]}]\right)
 + \quad \dots \\[8pt]
A_{\a\da} &=& \0A_{\a\da}  - \bt^j_\da \0\l_{j\a}
             - \bt^j_\da \bt^{i\db} \0\n_{\a\db} \0W_{ij} 
  \\[5pt]&&\quad
   - {1\over 3} \bt^j_\da \bt^{i\db}\bt^{l\dg}
\left(\0\n_{\a\db} \0\x_{lij\dg} - \e_{\db\dg}[ \0\l_{l} , \0W_{ij}] 
\right) \\[5pt]&&\quad
- {1\over 12} \bt^j_\da \bt^{i\db}\bt^{l\dg}\bt^{m\dd} 
         \0\n_{\a\db} \0g_{mlij\dg\dd} + \quad \dots \la{expa}\ea\ee 
It is easy to check that expansions \r{exp}, \r{expa}  satisfy the
$\D$-recursion relations \r{rec}-\r{recd}; and it is clear how higher
terms may be obtained from
the latter. The supersymmetry transformations of the component fields
\r{susy} may now be obtained immediately from the action of 
$ \d = \eb^{i\da} \der{\bt^{i\da}} + \eta^\a_i \der{\t^\a_i} $
on the non-chiral superfield $\widehat f_{ij}$ in \r{hatf} with its 
chiral \sf\ components $f_{ij} = 2 W_{ij}, f_{i\a}=\l_{i\a}, f_{\a\b}$ 
having the above the expansions.

The implication  $(iii)\Rightarrow (i)$, that the superconnection satisfies
the constraints (23) provided that the \eoms\ \r{comp} hold for the 
components may now be directly verified order by order in a \te\ by 
inserting \r{expa} in the constraints \r{csv} and \r{cvv}. An inductive 
proof of this implication is given in the next section. 

\vskip 10pt
\section{ The equivalence between \sf\ equations and component 
equations}
\vskip 10pt

In the previous sections we have obtained the proof of the implications
$(i)\Rightarrow (ii)\Rightarrow (iii)$ between our three sets of data.
The chain of the inverse implications,
$(iii)\Rightarrow (ii)\Rightarrow (i)$, and therefore the full 
equivalence between the three sets, may be proven by 
induction on the degree of $\bt$-homogeneity from the zero-order 
(component) equations, following the method of Harnad et al \c{hhls,hs}. 

The relations \r{rec}-\r{recd}  imply the following further $\D$-recursions:
\be\arr
&& \D f_{\da\db} = \D [ \n^\a_{\;\da}, \n_{\a\db} ] \  =\   
                 - \bt^i_{(\db} \n^\a_{\;\da)} \l_{i\a} \\[10pt]
&& \D (\n^\a_{\;\da} \l_{i\a})
 \  =\   \bt^j_\da \left( \{ \l^\a_j, \l_{i\a} \} + 2 \square W_{ij}\right)\ 
\\[10pt]
&& \D \left( \n^{\a\db} \n_{\a\db} W_{ij} - \{ \l^\a_i, \l_{j\a} \} 
\right)   \\[5pt]&&\quad =\    
 \bt^{k\db} \left( 2 [ \l^\a_k , \n_{\a\db} W_{ij} ] +
 \n^{\a\db} \n_{\a\db} \x_{ijk\db} + 2 [\n^\a_{\;\db}W_{k[i}, \l_{j]\a}]
 \right) \\[5pt] &&\quad  =\   
\bt^{k\db} \left( \n^{\a\db} \n_{\a\db} \x_{ijk\db} 
   - 2 [ \l_{[i}^\a , \n_{\a\da} W_{jk]} ]\right)\  \\[10pt]
&& \D \left( \n^{\a\da} \x_{ijk\da} -  [ \l^\a_{[i} , W_{jk]} ] 
\right)  \\[5pt]&&\quad  =\   
\bt^{k\db}\ \left(  \{ \l^\a_l , \x_{ijk\da} \} + 
   \n^{\a\da} g_{lijk\da\db} + \n^\a_{\;\db} [ W_{l[i} , W_{jk]} ]  
- 2 [\n^\a_{\;\db}W_{l[i}, W_{jk]}] 
\right.  \\[5pt]&&\qquad\qquad\qquad \left.
   -  \{ \l^\a_{[i} , \x_{jk]l\db} \}\right)  \\[5pt]&& \quad =\ 
\bt^{k\db} \left(  \n^{\a\da} g_{lijk\da\db }  
                  - J^\a_{lijk\da} \right) \la{eomsrec}\ea\ee  
and so on. In the general case we have
\be\arr&& \D \left( \n_\a^{\da_n} \phi_{i_{n+2}\dots i_1 \da_1\dots \da_{n} }
          -  J_{i_{n+2}\dots i_1 \a\da_1\dots \da_{n-1} }\right)
\\[5pt]&&\quad
=\  \bt^{i_{n+3}\da_{n}} \left( 
 \n_\a^{\da_{n+1}} \phi_{i_{n+3}\dots i_1 \da_1\dots \da_{n+1} } 
 - J_{i_{n+3}\dots i_1 \a\da_1\dots \da_{n} }\right)\la{eomrec}\ea\ee
Since these relations are linear in $\bt$ on the right, the $(n+1)$-st
order terms on the left are determined by the $n$-th order terms
of the coefficients of $\bt$ on the right. So if we assume that the 
\sf\ \eq s hold to order $n$ in $\bt$, the right-hand-sides of 
\r{eomsrec}, \r{eomrec}
vanish up to order $(n+1)$. Therefore, since the homogeneity operator 
is positive on the expressions above, it follows by induction that:

{\it The \sf\ \eq s hold if the (zeroth-order) component equations \r{comp}
are satisfied, i.e. $(iii)\Rightarrow (ii)$.} 

These relations also demonstrate that the entire tower of higher--spin
component \eq s are contained in the \sf\ \eq\ \r{ex} for $\x_{ijk\da}$,
or equivalently in the \sf\ \eq\ \r{eg} for $g_{ijkl\da\db}$.

We may similarly prove that the  $\D$--recursions \r{rec}-\r{recd}  
imply the relations (55). The first step of the induction follows
since at zero ($\bt$-independent) order, (55) are
manifestly implied by \r{rec}-\r{recd}. We now proceed to show that given the
$\D$--recursions \r{rec}-\r{recd}, the \sf\ relations (55) hold to 
order $(n+1)$ in $\bt$ provided they are valid to order $n$. 

Applying $\D$ to (55a) and using  the $\D$--recursions \r{rec}-\r{recd}, we obtain 
$$\arr
&&\D ( \n_{i\da} f_{\a\b}  - \half \n_{(\a\da} \l_{i\b)} ) \\[5pt]
&&\quad = - \n_{i\da} f_{\a\b} + 2 \bt^j_\da  [ W_{ij} , f_{\a\b} ]
+ \half \n_{i\da} (\bt^{j\db} \n_{(\a\db}\l_{j\b)}) \\[5pt]
&&\qquad + \half \bt^j_\da \{ \l_{j(\a},\l_{i\b)}\}
- \n_{(\a\da}\bt^{j\db}\n_{\b)\db}W_{ji} \ea$$
Therefore
$$\arr
&&(1+\D) ( \n_{i\da} f_{\a\b}  - \half \n_{(\a\da} \l_{i\b)} ) 
\quad\hbox{  (to order $(n+1)$ ) }\\[5pt]
&&\quad =\ 2 \bt^{j\db}  
    [ W_{ij} , \e_{\da\db}f_{\a\b} - [\n_{(\a\db},\n_{\b)\da}]]
\quad\hbox{ by the inductive hypothesis }\\[5pt]
&&\quad =\  0\ ,\ea$$ 
in virtue of \r{cvv1}, which holds to n-th order as a consequence of the 
inductive hypothesis. The relation (55a) therefore follows, since $(1+\D)$
is a positive-definite operator.

Similarly, applying  $\D$ to (55b) and using  the $\D$--recursions \r{rec}-\r{recd} 
yields 
$$\arr && \D ( \n_{i\da} \l_{j\a} - 2 \n_{\a\da} W_{ij} )\\[5pt] &&=\
 [[\D , \n_{i\da}], \l_{j\a}] 
+ 2 \n_{i\da} (\bt^{k\dg} \n_{\a\dg} W_{kj}) \\[5pt] &&=\quad
+ 2 \bt^k_\da  [ W_{ij} , \l_{k\a} ] 
- 2 \bt^{k\dg} \n_{\a\da}\x_{kij\dg}\ea$$
and using the inductive hypothesis, we obtain 
$$\arr
&&(1+\D )( \n_{i\da} \l_{j\a} - 2 \n_{\a\da} W_{ij} ) 
\quad\hbox{  (to order $(n+1)$ ) }\\[5pt]  &&\quad =\
-2\ \bt^k_\da \left(\n_\a^{\;\dg} \x_{ijk\dg } - [\l_{[i\a} , W_{jk]}] 
\right)\\[5pt]  &&\quad =\  0\  ,\ea$$
in virtue of \r{ex}, which, to n-th order, is a consequence of the 
inductive hypothesis.  

Now, applying $\D$ to (55c) and using  the $\D$--recursions \r{rec}-\r{recd}  yields 
$$\D ( \n_{i\da} W_{jk} - \x_{ijk \da})  \\[5pt]\quad =\
 - \n_{i\da} W_{jk} + 2\bt^l_\da [W_{il},W_{jk}] 
 + \n_{i\da}(\bt^{l\db}\x_{ljk\db}). $$
The inductive hypothesis therefore implies that
$$\arr&&(1+\D) ( \n_{i\da} W_{jk} - \x_{ijk \da})
 \quad\hbox{  (to order $(n+1)$ ) } \\[5pt]&&\quad =\
\bt^l_\da (2\ [W_{il}, W_{jk}]-[W_{i[l}, W_{jk]}]+[W_{l[i}, W_{jk]}])
\\[5pt]&&\quad\equiv\  0 ,\ea$$
yielding (55c). Note that this and the proofs of all further relations
in (55) follow from just the inductive hypothesis, whereas the proofs
of (55a,b) above require the satisfaction of the \sf\ \eoms\  \r{cvv1}
and \r{ex} to $n$-th order in $\bt$.

Similarly, for the next relation, the $\D$--recursions \r{rec}-\r{recd} 
imply that 
$$\arr&&(1+\D) (\n_{i\da} \x_{jkl\db} - g_{ijkl\da\db} 
                  - \e_{\da\db} [ W_{i[j}, W_{kl]}] )
 \quad\hbox{  (to order $(n+1)$ ) } \\[5pt]&& =\
2\bt^m_\da [W_{im}, \x_{jkl\db}] 
 - \bt^{m\dg} \n_{i\da} \left( g_{ijkl\da\db} 
                  + \e_{\da\db} [ W_{i[j}, W_{kl]}] \right)
- \bt^{m\dg} \psi_{mijkl\dg\da\db} \\[5pt]&&\quad 
+ \bt^m_{(\da} \left( {2\over 3} [ W_{m[i} , \x_{jkl]\db)} ]
	  - {1\over 3} [ W_{[ij} , \x_{kl]m\db)} ]  \right) 
-  \bt^m_{[\da} 
     \left( [ \x_{mi[j\db]} , W_{kl]}] - [W_{i[j},\x_{kl]m\db]} ] \right)
\\[5pt]&& =\ 0 \ea$$
if \r{bh} is assumed to hold to order $n$. 
All remaining relations in (55) follow similarly
from the positive-definiteness of $(1+\D)$ and the fact that 
$$\arr&& 
(1+\D) \left( \n_{i_{n+2}\da_1} \phi_{i_{n+1}\dots i_1 \da_2\dots \da_{n}}
 - \phi_{i_{n+2}\dots i_1 \da_1\dots \da_{n}}  
\right.\\[5pt]&&\qquad\qquad \left. 
 - \e_{\da_1(\da_2} \Z_{i_{n+2}\dots i_1\da_3\dots \da_{n})} \right)
\qquad\hbox{ (to order $(n+1)$ ) } \\[5pt]&&\quad =\  0
 \quad\hbox{  by the inductive hypothesis.}\ea$$

We may now proceed to prove by induction that the $\D$--recursions 
\r{rec}-\r{recd}, together with (55) to order $n$, imply the constraint 
equations \r{con} to order $(n+1)$.
The zero-th order relations are manifest in virtue of the recursion 
relations \r{reca} and the definition of the curvature $f_{\a\b}$.
The rest follows by action of $\D$ on the  constraints \r{con}. Thus
$$\arr &&\D \left(\{\n_{i \da}, \n_{j \db}\} - 2\e_{\da \db} W_{i j}\right)
\\[5pt]&&\quad
=\ -2 \{\n_{i \da}, \n_{j \db}\} + 2 \n_{i \da} ( \bt_\db^k W_{jk}) 
+ 2 \n_{j \db}( \bt_\da^k W_{ik}) 
- 2 \e_{\da \db}\bt^{k\dg} \x_{ijk\dg}\  .\ea$$
Assuming (55c) to order $n$, we therefore have that
$$ (2+\D) \left(\{\n_{i \da}, \n_{j \db}\} - 2\e_{\da \db} W_{i j}\right)
 \quad\hbox{ (to order $(n+1)$ )\quad }=\ 0\  ,$$
from which the first equation in \r{con} follows since $(2+\D)$ 
is a positive operator. Similarly, \r{rec}-\r{recd} imply that 
$$\D \left([\n_{i \da}, \n_{\b \db}] - \e_{\da \db}\l_{i \b}\right)
\\[5pt]=\ - [\n_{i \da}, \n_{\b \db}] - 2 \bt^k_\da \n_{\b\db} W_{ik} 
  - \n_{i \da}(\bt^k_\db \l_{k\b}) - 2 \bt^k_{[\da} \n_{\b\db]} W_{ki} $$
and assuming the validity of (55b) to order $n$, we have the relation
$$(1+\D) \left([\n_{i \da}, \n_{\b \db}] - \e_{\da \db}\l_{i \b}\right)
 \quad\hbox{ (to order $(n+1)$ )\quad }  =\ 0\ ,$$
which in turn implies the second equation in \r{con}.
Finally,
$$\arr&&\D\left([\n_{\a \da}, \n_{\b \db}] - \e_{\da \db}  f_{\a \b}\right)
 \quad\hbox{ (to order $(n+1)$ )\quad }\\[5pt]&&\quad=\ 
\bt^i_\da \n_{\b\db} \l_{i\a} - \bt^i_\db \n_{\a\da} \l_{i\b}
- \half \e_{\da \db} \bt^{i\dg}\n_{(\a \dg}\l_{i\b)}
\\[5pt]&&\quad =\ 0\  \ea$$
if the \eom\ for $\l_{i\a}$ are assumed to hold to order $n$, since
$\n_{\b\db} \l_{i\a} = \half \n_{(\b\db} \l_{i\a)}$ in virtue of \r{el}.

This completes the proof of the chain of implications   
$(iii)\Rightarrow (ii)\Rightarrow (i)$. We have therefore
demonstrated the full equivalence between our three sets of data.

\vskip 10pt
\section{ The conserved tensor supercurrents}
\vskip 10pt

Gauge--invariance of the $\pmatrix{ N\cr 4 }$ \sf\  functionals \r{l4} 
yields this number of second rank traceless (i.e. satisfying
$\e^{\a\b}\e^{\da\db} T_{ijkl\a\da ,\b\db} =0 $) conserved \sf\ tensors
having the form
\be \arr
T_{ijkl\a\da ,\b\db}  &=& \Tr \left(
     g_{ijkl\da\db} f_{\a\b}  +  \n_{\a\db} \l_{[i\b} \x_{jkl]\da}
   -    \l_{[i\a} \n_{\b\da} \x_{jkl]\db}          \right.\\[8pt]
 && \qquad  + {1\over 2} \l_{[i\b} \n_{\a\da} \x_{jkl]\db}
        - {1\over 2} \n_{\a\da} \l_{[i\b} \x_{jkl]\db}
 + {2\over 3} \e_{\db\da}\e_{\b\a} \{ \l^\g_{[i}, \l_{j\g}\} W_{kl]} 
\\[8pt] && \qquad  
     - {1\over 3} \n_{(\a\da} W_{[ij}\n_{\b)\db} W_{kl]}
             + {1\over 3} W_{[ij} \n_{\a\da}\n_{\b\db} W_{kl]} 
\left.\right)  .\la{tr} \ea\ee
The $\Tr$ in these expressions denotes the gauge algebra trace.
The expression \r{tr} is in fact the unique traceless linear combination
of the three existing second rank conserved tensors,
\be\arr
T^{(1)}_{ijkl\a\da ,\b\db}&=& \Tr \left(
   g_{ijkl\da\db} f_{\a\b} -  \l_{[i\a} \n_{\b\da} \x_{jkl]\db}
   -  \n_{\b\da} W_{[ij}\n_{\a\db} W_{kl]}  \right) \\[8pt]
T^{(2)}_{ijkl\a\da ,\b\db} &=& \Tr \left(
 {1\over 2} \l_{[i\b} \n_{\a\da} \x_{jkl]\db}
 - {1\over 2} \n_{\a\da} \l_{[i\b} \x_{jkl]\db}
 +  \n_{\a\db} \l_{[i\b} \x_{jkl]\da} \right. \\[5pt] &&\qquad\left.
 +  \e_{\b\a} \e_{\db\da} \{ \l^\g_{[i}, \l_{j\g} \} W_{kl]}  \right) 
 \\[8pt]
T^{(3)}_{ijkl\a\da ,\b\db} &=& \Tr {1\over 3}\left(
    W_{[ij} \n_{\a\da}\n_{\b\db} W_{kl]}
    -  \n_{\a\da} W_{[ij} \n_{\b\db} W_{kl]}
    + 2 \n_{\b\da} W_{[ij} \n_{\a\db} W_{kl]}  \right.  
\\[5pt]  &&\qquad\left.
    - \e_{\b\a} \e_{\db\da} \{ \l^\g_{[i}, \l_{j\g} \} W_{kl]}\right)\ 
  .\la{stresses}\ea\ee
The conservation of these tensors is a non-trivial consequence of the
\sf\ \eoms. Thus, using the \eom\ \r{eg}, together with the operator
identity
\be \n^{\a\da}\n_{\b\da} = \d^\a_\b \square + f^\a_\b  \la{nn}\ee
and the cyclic property of the trace, we obtain
$$\arr \p^{\a\da} T^{(1)}_{ijkl\a\da ,\b\db}&=& \Tr \n^{\a\da} \left(
   g_{ijkl\da\db} f_{\a\b} -  \l_{[i\a} \n_{\b\da} \x_{jkl]\db}
   -  \n_{\b\da} W_{[ij}\n_{\a\db} W_{kl]}  \right) \\[5pt]
&=& \Tr  \left( -  \l_{[i\b} \square \x_{jkl]\db}
   -  2 \square W_{[ij}\n_{\b\db} W_{kl]}  \right) \\[5pt]
&=& 0\ \hbox{\quad in virtue of \r{ex} and \r{wx}.}\ea$$
Similarly,
$$\arr \p^{\a\da}T^{(2)}_{ijkl\a\da ,\b\db} &=& \Tr \left(
\l_{[i\b} \square \x_{jkl]\db} + \n_{\a\db}\l_{[i\b}\n^{\a\da}\x_{jkl]\da}
+ \n_{\b\db}( \{ \l^\g_{[i}, \l_{j\g} \} W_{kl]} ) \right)\\[5pt]
&=& 0 \ea$$
also in virtue of \r{ex} and \r{wx}. Finally \r{nn} and the further 
operator identity
\be \n^{\a\da}\n_{\a\da}\n_{\b\db} = 2\n_{\b\db}\square +
                                    2 f^\a_\b \n_{\a\db} \la{nnn}\ee
implies that
$$\arr \p^{\a\da} T^{(3)}_{ijkl\a\da ,\b\db} &=& {2\over 3}\Tr\left(
2 W_{[ij}\n_{\b\db} \square W_{kl]} +2 \square W_{[ij} \n_{\b\db} W_{kl]}
  -\n_{\b\db}( \{ \l^\g_{[i}, \l_{j\g} \} W_{kl]} ) \right)\\[5pt]
&=& 0\ \hbox{\quad in virtue of \r{ew}.}\ea$$

The gauge--invariant tensors \r{stresses} have conserved superpartners. 
The lower rank conserved spin--tensors are
\be\begin{array}{lll}
T_{ijk\a\da ,\b} &=& \Tr  (2 f_{\a\b} \x_{ijk\da}
   - \n_{\a\da} \l_{[i\b} W_{jk]}
   +  \l_{[i\b} \n_{\a\da} W_{jk]}
   - 2 \l_{[i\a} \n_{\b\da} W_{jk]} )   \\[5pt]
T^{(1)}_{ijklm\a\da ,\db} &=& \Tr (
     4 \l_{i\a} g_{jklm\da\db} - \l_{[j} g_{klm]i\da\db}
     - 4 \x_{i[jk\da} \n_{\a\db} W_{lm]}           \\[5pt]   &&\qquad
     + 6 \n_{\a\db} W_{i[j} \x_{klm]\da}
   - 5 \e_{\da\db} W_{i[j} [ \l_{k\a} ,W_{lm]} ]   ) \\[5pt]
T^{(2)}_{ijklm\a\da ,\db} &=& \Tr (
 \n_{\a\da}W_{i[j}\x_{klm]\db} -  W_{i[j}\n_{\a\da}\x_{klm]\db}
	    - 2 \n_{\a\db} W_{i[j}\x_{klm]\da} \\[5pt]   && \qquad
   + 2 \e_{\da\db} W_{i[j} [ \l_{k\a} ,W_{lm]} ]   ) \\[5pt]
T^{(3)}_{ijklm\a\da ,\db} &=& \Tr (
 \n_{\a\da}\x_{i[jk\db} W_{lm]} -  \x_{i[jk\db} \n_{\a\da} W_{lm]}
	 + 2 \x_{i[jk\da} \n_{\a\db} W_{lm]} \\[5pt]   &&\qquad
   + 2 \e_{\da\db} W_{i[j} [ \l_{k\a} ,W_{lm]} ]   ) \\[5pt]
T_{ijkl\a\da} &=& \Tr  ( 3 \l_{i\a} \x_{jkl\da} + \l_{[j\a} \x_{kl]i\da}
          +  2 \n_{\a\da} W_{i[j} W_{kl]} - 2 W_{i[j}\n_{\a\da} W_{kl]}  )\
	                      .\ea\ee
All these tensors satisfy the conservation law
$$ \p^{\a\da} T_{i\dots m \a\da , \dots} =\  0   $$
in virtue of the equations of motion, for instance,
$$ \arr \p^{\a\da} T_{ijkl\a\da} &=& \Tr \n^{\a\da} 
  \left( 3 \l_{i\a} \x_{jkl\da} + \l_{[j\a} \x_{kl]i\da}
 +  2 \n_{\a\da} W_{i[j} W_{kl]} - 2 W_{i[j}\n_{\a\da} W_{kl]} \right)
\\[5pt]&=& \Tr\left( 3 \l_{i\a} [\l^\a_{[j}, W_{kl}] 
 + 2 \l_{[j\a} [\l^\a_{[k}, W_{l]i}] + \l^\a_i [\l_{[j\a}, W_{kl}] 
\right. \\[5pt]&&\qquad \left.
 + 2 \{ \l^{i\a}, \l_{[j\a} \} W_{kl]} -2 W_{i[j} \{ \l^\a_k, \l_{l]\a}\}
\right)\\[5pt] &=& 0\  .\ea$$
These may be used to couple $N\ge 4$ \sd gauge theories to gravity and 
supergravity, whereas there are no appropriate conservation laws for
such couplings of $N\le 3$ \sd theories.

\vskip 10pt
\section{ Concluding remarks}
\vskip 10pt

We have demonstrated that the \sdyme afford supersymmetrisation beyond 
the conventionally `maximal' \N4 extension and thus yield non-trivial 
four--dimensional Lorentz covariant systems of equations invariant
under N-extended rigid Poincar\'e supersymmetry for arbitrary values of $N$.

The \sdy constraints for the supercurvature have been shown to imply 
the existence of \sf s of arbitrarily high spin and we have also 
demonstrated the complete equivalence of these constraints to the 
component \eoms , which for $N\ge 4$  provide possibly the unique 
consistently coupled realisations of the zero rest--mass Dirac--Fierz 
\eq s for arbitrary spin fields. The consistency of our systems is
actually a consequence of the {\it matreoshka phenomenon}: the 
$N$-extended system nestles within the $(N+1)$-extended system completely
intact. The extra fields of the latter have interactions governed by
a source current which depends only on lower spin fields and which
does not require modification on further supersymmetrisation. 

We have further demonstrated that the $N\ge 4$ systems of equations imply 
the conservation of a stress tensor, and \susy\ generalisations. They
may therefore be coupled nontrivially to the Einstein equations
and to supergravity. This is unlike the $N\le 3$ \sdym theories, which 
have no appropriate stress--like tensors.

Our \N5 theory, moreover, is probably the unique supersymmetric theory 
in which a spin ${3\over 2}$ field is coupled to a vector field, without 
requiring a spin 2 coupling as well for consistency \c{os}. However,
our systems also allow {\it locally} \susy\ generalisations, i.e. 
arbitrary--$N$ \sd supergravities. These have spin $1$, spin ${3\over 2}$ 
and spin $2$ gauge--invariances, as well as both Yang-Mills and 
gravitational coupling constants. In chiral \ss these arbitrarily extended
\sd supergravity \eq s  take the form (23) as well, with the covariant
derivatives being {\it generally covariant} ones in chiral \ss \c{sdsg}. 
The systematic unravelling of these constraints requires a generalisation 
of the procedure presented in this paper. The appropriate generalisation 
of the $\D$-gauge, which eliminates all $\bt$-dependence of diffeomorphism
as well as gauge--transformation parameters has been discussed by us
in \c{buckow}.

Our \ssd systems provide an infinitely large enhancement of an
already rich class of conformally invariant exactly soluble systems in 
four dimensions (the \N0 \sdy equations), a \susy version of the 
twistor transform providing a method of constructing explicit 
solutions \c{matr,pl}.
In view of recent discussions about the centrality of the \sdyme and their
twistor transform in the theory of integrable systems (e.g. \c{ms}), the
significance of these extensions seems obvious. Reductions are likely
to yield all possible integrable couplings of the lower dimensional
systems hitherto found to descend from the \sdym \eq s. Our systems indeed 
provide a non-trivial {\it \sdym hierarchy} of integrable systems. It 
seems likely that the recently discussed large--$N$ extensions of the KdV 
equations \c{g} are reductions of our systems.

\vskip 20pt\noindent
{\Large\bf In memoriam  V. I. Ogievetsky (1928  -- 1996)}
\vskip 15pt
\noindent
Victor Isaakovich Ogievetsky died on 23rd March, 1996. 
The work reported in this paper was largely completed in the autumn 
of 1995 and Victor Isaakovich worked courageously and with unfailing 
enthusiasm on this text during the difficult last months of his life.
The memory of an imaginative physicist, an inspiring teacher and a 
generous soul will always endure.
\goodbreak
\goodbreak
\end{document}